# Novel Fourier-domain constraint for fast phase retrieval in coherent diffraction imaging

Tatiana Latychevskaia*, Jean-Nicolas Longchamp, and Hans-Werner Fink

Physics Institute, University of Zurich
Winterthurerstrasse 190
8057 Zurich
Switzerland

Corresponding author:
e-mail: tatiana@physik.uzh.ch

## Abstract
Coherent diffraction imaging (CDI) for visualizing objects at atomic resolution has been realized as a promising tool for imaging single molecules. Drawbacks of CDI are associated with the difficulty of the numerical phase retrieval from experimental diffraction patterns; a fact which stimulated search for better numerical methods and alternative experimental techniques. Common phase retrieval methods are based on iterative procedures which propagate the complex-valued wave between object and detector plane. Constraints in both, the object and the detector plane are applied. While the constraint in the detector plane employed in most phase retrieval methods requires the amplitude of the complex wave to be equal to the squared root of the measured intensity, we propose a novel Fourier-domain constraint, based on an analogy to holography. Our method allows achieving a low-resolution reconstruction already in the first step followed by a high-resolution reconstruction after further steps. In comparison to conventional schemes this Fourier-domain constraint results in a fast and reliable convergence of the iterative reconstruction process.

## 1. Introduction

The idea of recovering the structure of non-crystalline objects from their far-field diffraction patterns dates back to 1952, when Sayre wrote a quarter-of-page commentary to Shannon's theorem [1]. In 1972, Gerchberg and Saxton [2] suggested an iterative scheme which established the foundation for phase retrieval methods applied in CDI [3]. Along with the improvements of experimental techniques to obtain coherent diffraction patterns, such as the development of x-ray free-electron lasers [4], numerical routines for phase recovery have also been perfected [5].

Although the object shape cannot easily be recognized in its far-field image, it can be uncovered by iterative reconstruction processes. As in the original iterative scheme, all reconstruction routines derived thereafter [6-8] require some information about the object to be known beforehand. For instance, the object transmission function must be real and positive, as in the case of x-ray scattering at the electron density of the molecules, and the object must occupy some pre-defined area in space. The latter allows applying a "support" with known transmission properties [9]. Despite these constraints, the iterative routines do not always converge, and even if they do, different iterative runs might lead to different results. Usually hundreds of such runs are required and their results are subsequently averaged [10-12]. The difficulty of the numerical phase retrieval from experimental diffraction patterns also triggered a number of novel experimental techniques [13-15].

While there is an intensive search for optimal constraints in the object plane to improve the reconstruction routines, the possibility to apply constraints in the Fourier domain has not yet fully been explored. As in the original phase retrieval routine [2], most reconstruction methods employ the constraint in the detector plane that the amplitude of the complex-valued wavefront must be equal to the square root of the measured intensity. Only a few recent studies consider using other constraints in the Fourier domain: Dilanian et al. [16] applied a statistically regularized constraint to account for noise and Zhang et al. [17] varied the measured intensity constraint as $I^\gamma$, and noticed that if $\gamma>0.5$ the iterative routine converges much faster. Below we present a novel constraint in the Fourier-domain which exhibits two major advantages: it allows dropping the requirement of a well-defined support around the object and it provides fast, reliable and unambiguous reconstructions.

## 2. Theoretical background

In a diffraction pattern, the measured intensity can be written as $I_0=U_0U_0^*$, where $U_0$ is the complex-valued wave in the detector plane. The iterative reconstruction process based on the Gerchberg-Saxton algorithm [2] is schematically illustrated in Fig. 1. As in common phase retrieval algorithms, the iterative loop in our method starts with the complex-valued field in the detector plane $U_1$: its amplitude is given by the square root of the measured intensity, and its phase is chosen to be randomly distributed. In the iterative process $U_i'$, i=1,2,... is the updated complex-valued wavefront in the detector plane after each iteration. In common phase retrieval methods, the phase of the iterated field is adapted for the next iterative loop, while the amplitude is replaced by the square root of the measured intensity:

$$U_{i+1} = \frac{U_i'}{|U_i'|}\sqrt{I_0} \qquad (1)$$

We propose using the following constraint instead:

$$U_{i+1} = \frac{U_i'}{|U_i'|^2 + \varepsilon}U_0U_0^* = L_iU_0U_0^* \qquad (2)$$

This formula is derived via a direct analogy to holography, where the transmission of a recorded hologram $H$ is given by $H\sim R^*O+RO^*$. Here, $R$ denotes the reference wave and $O$ the object wave. In the reconstruction process, the hologram is multiplied with the reference wave resulting in the straightforward reconstructed object wave $RH/|R|^2$. This approach can be applied for reconstructing a diffraction pattern if we formally treat the measured intensity of the diffraction pattern $U_0U_0^*$ as a hologram $H$ and $U_i'$ as the reference wave $R$, see Eq. (2). This novel constraint unlike the one given by Eq. (1), does not only account for the phase of $U_i'$, but also for its amplitude. The latter must gradually approach the amplitude of $U_0$. In order to avoid division by zero, a constant $\varepsilon$ is added in the denominator. For measured intensities of up to $10^7$ counts per pixel, $\varepsilon$ is selected in the range of 0.1-1000. In Eq. (2), $L_i$ denotes the Fourier-domain filter, a complex-distributed function altered at each iteration. Any additional constraint in the object plane can still be selected independently of the Fourier-domain constraint.

## 3. Simulation results

The performance of this novel constraint is demonstrated by the reconstruction of simulated diffraction patterns of some test objects. We begin with a binary object in form of the letter "β", the results are shown in Fig. 2. To mimic experimental conditions we added a noise distribution to the simulated diffraction pattern:

noise(m,n)=(signal(m,n)/SNR)*random(-0.5;0.5) [3], where SNR=5 is the assumed signal-to-noise ratio. The random function generates numbers in the range (-0.5;0.5) at pixel position (m,n) with 500x500 pixels in total. The constraint in the object plane implies that the transmission function is real and positive; the phase is thus set to 0. It is important to note that no support masks are applied in the object domain.

To achieve a fast and reliable convergence of the algorithm, the reconstruction is done in several steps using a modified constraint:

$$U_{i+1} = \frac{U_i{'}}{|U_i{'}|^2 + \varepsilon} \mu U_0 U_0^*. \qquad (3)$$

Here, the parameter $\mu$ is stepwise adjusted to 1. At the first iterative run (step one) $\mu_1$ is set to some value $0<\mu_1<<1$, and the solution approaches $U_i^{(1)} \to \sqrt{\mu_1} U_0$. This modified $L$-filter

$$L_i^{(1)} \to \frac{\sqrt{\mu_1} U_0}{\mu_1 |U_0|^2 + \varepsilon}, \qquad (4)$$

acts as a band-pass filter for the measured intensity distribution, as illustrated in Fig. 3. For the 1$^{st}$ step $\mu$ is selected such that the filter acts as a low-pass filter. For the first 90 iterations $\mu$ was set to $\mu_1$=0.001 and $\varepsilon_1$=1. The result of this iterative run is a reconstruction of a low-resolution image of the object, presented in Fig. 2(c).

The error function, calculated as

$$\text{Error} = \frac{\sum \left\| |U_0| - |U_i| \right\|}{\sum |U_0|}, \qquad (5)$$

where $|U_0|$ is the square root of the measured intensity and $|U_i{'}|$ is the updated amplitude of the complex wave at the detector plane after each iteration, shows fluctuations around some constant value, as evident from the plot in Fig. 2(e). The obtained in the first step complex-valued distribution $\sqrt{\mu_1} U_i^{(1)}$ is then used as the input for the second iterative run (step two), where $\mu_2$ is set to $\mu_2$=1, and thus $U_i^{(2)} \to U_0$. The result of the second iterative run (step two), presented in Fig. 2(d), where $\mu_2$=1 and $\varepsilon_2$=1 were used, required just less than 30 iterations to refine the low-resolution reconstruction. The error function quickly decreases, as shown in Fig. 2(e) and reaches 0.25 after a total of 120 iterations. The iterative process was terminated once the error reached a stable minimum. Alternatively, the low-resolution result obtained from the first iterative run can be used as an initial distribution for any other phase retrieval method.

It is worth noting that the method presented here has an advantage comparing to the conventional CDI reconstruction routines that it guarantees a quick low-resolution reconstruction already after one step. The low-resolution reconstruction can then be refined in a subsequent step towards the high-resolution reconstruction.

### 3.1 Simulated diffraction pattern of a ribosome

Next, we simulate and reconstruct a diffraction pattern of a non-binary object, a ribosome, see Fig. 4. The electron density map of a ribosome S70 from E. coli has been obtained by cryo-electron microscopy by averaging over 52'181 molecules (www.pdb.org) [18]. A simulated diffraction pattern of an individual ribosome is shown in Fig. 4(b). It was obtained by simulating an X-FEL pulse of $10^{12}$ photons with a wavelength of $\lambda$=1.4 Å that was focused onto a 100 nm region. A low-resolution reconstruction was obtained after 105 iterations in the 1$^{st}$ step, using $\mu_1$=0.0001 and $\varepsilon_1$=10. The final high-resolution reconstruction was obtained in the 2$^{nd}$ step after 101 iterations with $\mu_2$=1 and $\varepsilon_2$=1000. A total of 206 iterations were required and the final error amounts to 0.25. The reconstructed electron density, see Fig. 4(d), is in good agreement with the original object, Fig. 4(a), taking into account shift and flip invariance.

### 3.2 Diffraction pattern of a particle field with missing information in the central region

In the following example we apply our method to a simulated diffraction pattern of differently-sized individual spheres ranging from 14λ to 22λ in diameter and being distributed in space in such a way that their surfaces are separated by at least 24λ, see Fig. 5. This simulation relates to real experiments, like imaging a dilute micrometer-sized particle distribution with visible light or nanometer-sized clusters with X-rays or electrons. The intensity in the central region of the diffraction pattern is set to zero to mimic common experimental conditions where the central overexposed area on a detector is blocked and thus the low-resolution signal missing.

The diffraction pattern created by just one of the individual scatterer is spread over a large angular regime. While blocking out the central part of the diffraction pattern, sufficient information about this one scatterer is still contained in the higher order interference fringes. It turns out that this is also true for a set of individual scatterers, provided they are sufficiently distant from one another. Under these conditions, the positions of all individual objects can be recovered from a diffraction pattern even without the unavailable information in the central area of the diffraction pattern.

### 3.3 Diffraction pattern of a phase-shifting object

In this sub-section we simulate and reconstruct a diffraction pattern of a phase-shifting object, which is considered to be a difficult object for reconstruction. A set of round objects with constant phase shifts between 0 and π is selected as a test object, see Fig. 6. The simulated diffraction pattern was reconstructed by our method and, for comparison, by the shrinkwarp method [7]. In the reconstruction obtained by our method, the parameters $\mu_1$=0.3 and $\varepsilon_1$=1000 were chosen for the 1$^{st}$ step, and the constraint in the object plane is that the phase of the transmission function is set to 0. This allows obtaining a low-resolution reconstruction of the object after only 48 iterations. In the 2$^{nd}$ step, $\mu_2$=1 and $\varepsilon_2$=10000 are selected and the phase in the object domain is free from any constraint at those pixels where the amplitude exceeds 30% of the maximum of the amplitude, and it is set to 0 at the other pixels. In the 2$^{nd}$ step, 43 iterations are required to retrieve both, the amplitude and the phase distribution of the object, as illustrated in Fig. 6 (c)-(d). The evolution of the error is displayed in Fig. 6(g), and reaches a final value of 0.40. The very same diffraction pattern was also reconstructed employing the shrinkwrap method (Fig. 6(e)-(f)), using the hybrid input-output iterative scheme with the feedback parameter β=0.9. The supporting masks were created at 5% threshold of the maximum of the amplitude in the object plane and convoluted with a Gaussian (σ=3). The supporting mask was updated every 20 iterations. A total of 750 iterations were necessary to retrieve the complex-valued object with a final error of 0.47. The evolution of the error is shown in Fig. 6(h).

### 4. Experimental results

We also tested our novel reconstruction technique against experimental diffraction images. As sample, a non-transparent gold covered glass plate was used in which the letter "ℏ" about 90 μm in height was engraved by means of a focused ion beam into a 15 nm thick layer of Au deposited onto a glass slide. The sample was placed into a 532 nm wavelength laser beam and the diffraction pattern was recorded at a 135 mm distant screen by a CCD camera. The diffraction pattern within an area of 90x90 mm$^2$ is sampled with 500x500 pixels. The recorded diffraction pattern was centered by the symmetry of its peaks and the noise level of the camera (about 50 counts per pixel) was subtracted. Finally, taking the modulus of the resulting diffraction pattern allows the elimination of small negative values due to noise subtraction. The reconstruction was accomplished in three steps with the results displayed in Fig. 7. In all three steps the constraint in the object plane was that the phase of the retrieved transmission function was set to zero; no other constraints such as masks or "supports" were applied. In step one, 200 iterations were done using $\varepsilon_1=10$ and $\mu_1=0.0001$. In step two, 360 iterations were done using $\varepsilon_2=5000$ and $\mu_2=1$. Finally, in step three, 40 iterations were performed using $\varepsilon_3=500$ and $\mu_3=1$ to decrease the value of $\varepsilon$. Still, this third and last step resulted in only a minor improvement of the reconstruction, and could as well be skipped. By applying a larger number of iterations, the reconstructed image remained unchanged except for slight oscillations in the noise distribution. Thus, the stable reconstruction was already achieved after a total of 600 iterations. For comparison, we also tried to employ the "shrink-wrap" algorithm to the experimental diffraction pattern; however it failed to achieve a stable reconstruction resembling the object. The reason could be that the "shrink-wrap" algorithm relies on a continuously updated support mask, which could not be well defined for our experimental sample as there were some fine scratches around the object as verified by optical microscopy. Instead, we employed a reconstruction method using a support in the form of a fixed tight mask. In this case, about 4000 iterations were needed in a successfully convergent run to achieve a presentable reconstruction, as shown in Fig. 7(f).

## 5. Conclusion

In summary, we invented a Fourier-domain filter for fast and reliable object retrieval from diffraction images. The reconstruction method using this novel constraint does not require a "support" or a "mask". Thus, in contrast to all other reconstruction routines, any initial guess about the shape of the object is irrelevant for the convergence of our routine. This independence of the object morphology shall be highly beneficial for the reconstruction of objects that cannot be confined within a certain area suitable for "masking", or for objects that are surrounded by some features which cause unwanted contributions to the diffraction pattern. A low-resolution reconstruction of the object is already achieved after just a few iterations. This low-resolution reconstruction can then be either further refined by additional iterations or it can be used as an input for any other conventional phase retrieval method. While we have demonstrated our method using light optical diffraction here, it can of course be applied to data obtained with any other coherent radiation, be it X-rays [5] or electrons [19-20].

## Acknowledgements

We would like to thank Elvira Steinwand for the sample preparation and Mirna Saliba for careful reading of the manuscript. The Forschungskredit of the University of Zurich and the Swiss National Science Foundation are gratefully acknowledged for their financial support.

**Figures**

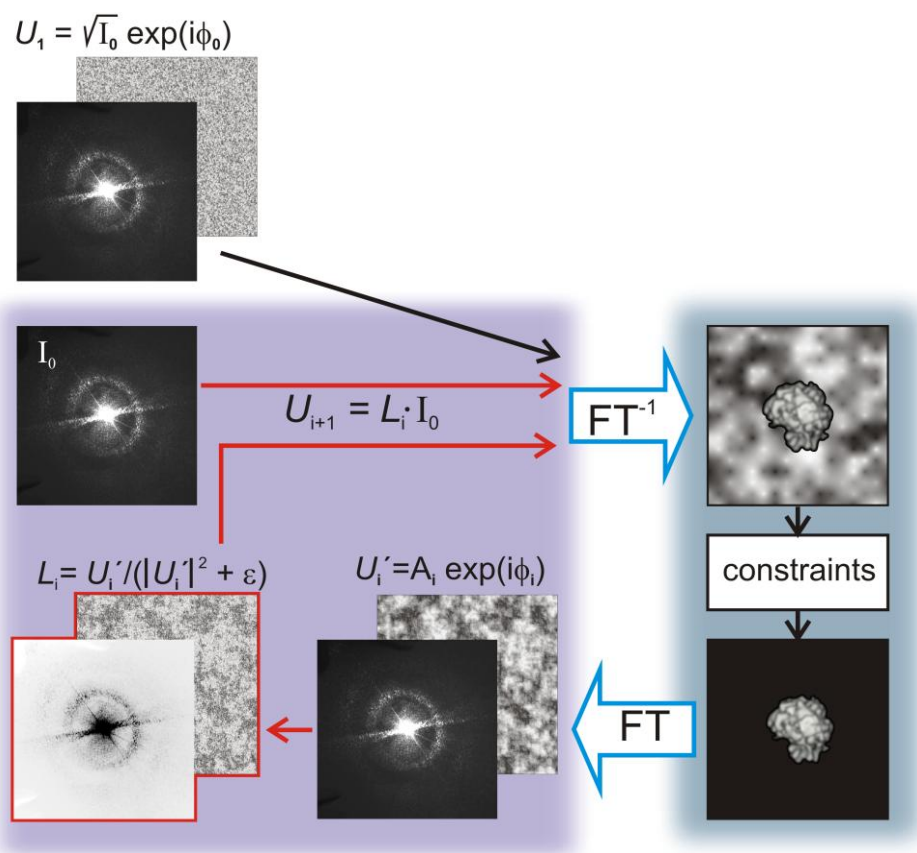

Fig. 1. Schematic representation of the iterative recovery of an object from its coherent diffraction pattern. The violet color highlights the Fourier domain while the blue color highlights the object domain. The red lines emphasize the filter components. The forward Fourier transform of the updated object distribution results in the complex-valued $U_i'$ which is used to build the Fourier-domain filter $L_i$. The product of $L_i$ and the measured intensity $I_0$ provides $U_{i+1}$, the input for the next iteration.

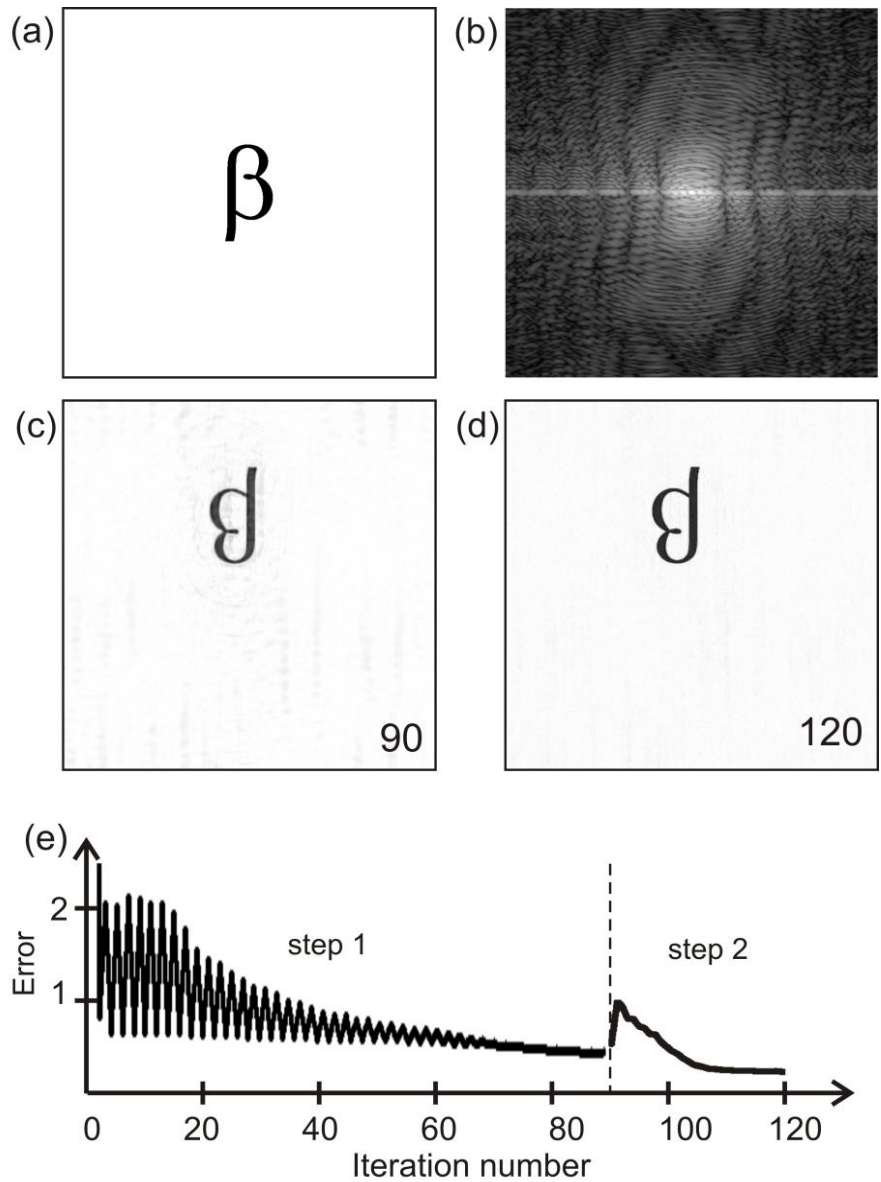

Fig. 2. Two-step recovery of the letter β from its simulated diffraction pattern. (a) Inverted amplitude of the transmission function of the test object. (b) Simulated diffraction pattern shown in logarithmic intensity scale. (c) Inverted amplitude of the low-resolution reconstruction of the object obtained after the first step consisting of 90 iterations with $\mu_1=0.001$ and $\varepsilon_1=1$. (d) Inverted amplitude of the object after the second step using additional 30 iterations with $\mu_2=1$ and $\varepsilon_2=1$. (e) Error as a function of iteration number.

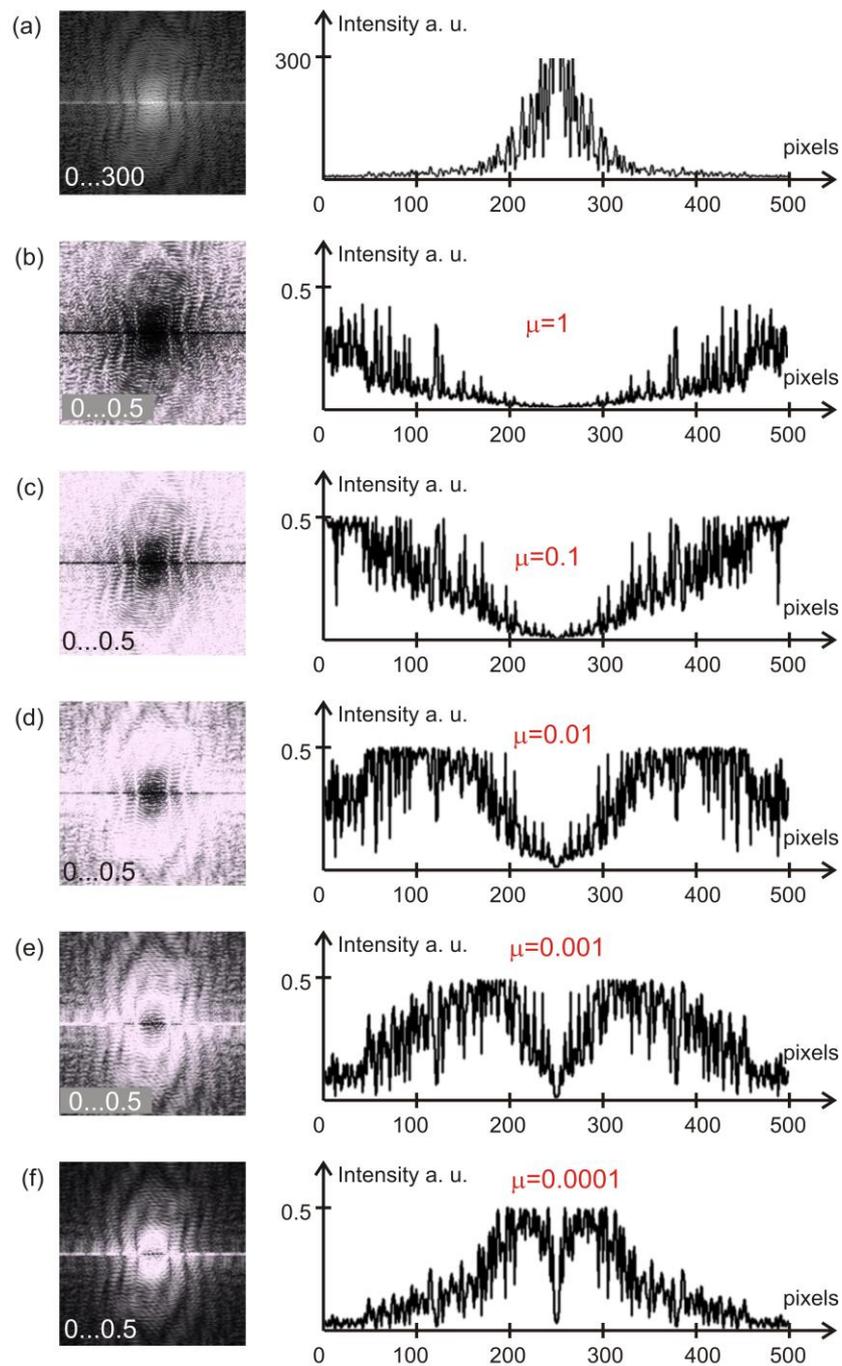

Fig. 3. Amplitude distributions of the *L*-filter for different values of the parameter µ. (a) Amplitude of the simulated diffraction pattern of "β" and its intensity profile along the middle horizontal line (the maximum of the intensity is cropped). (b)-(f) Left column: amplitude distributions of the *L*-filter. Right column: intensity profiles of the *L*-filter along the middle horizontal line.

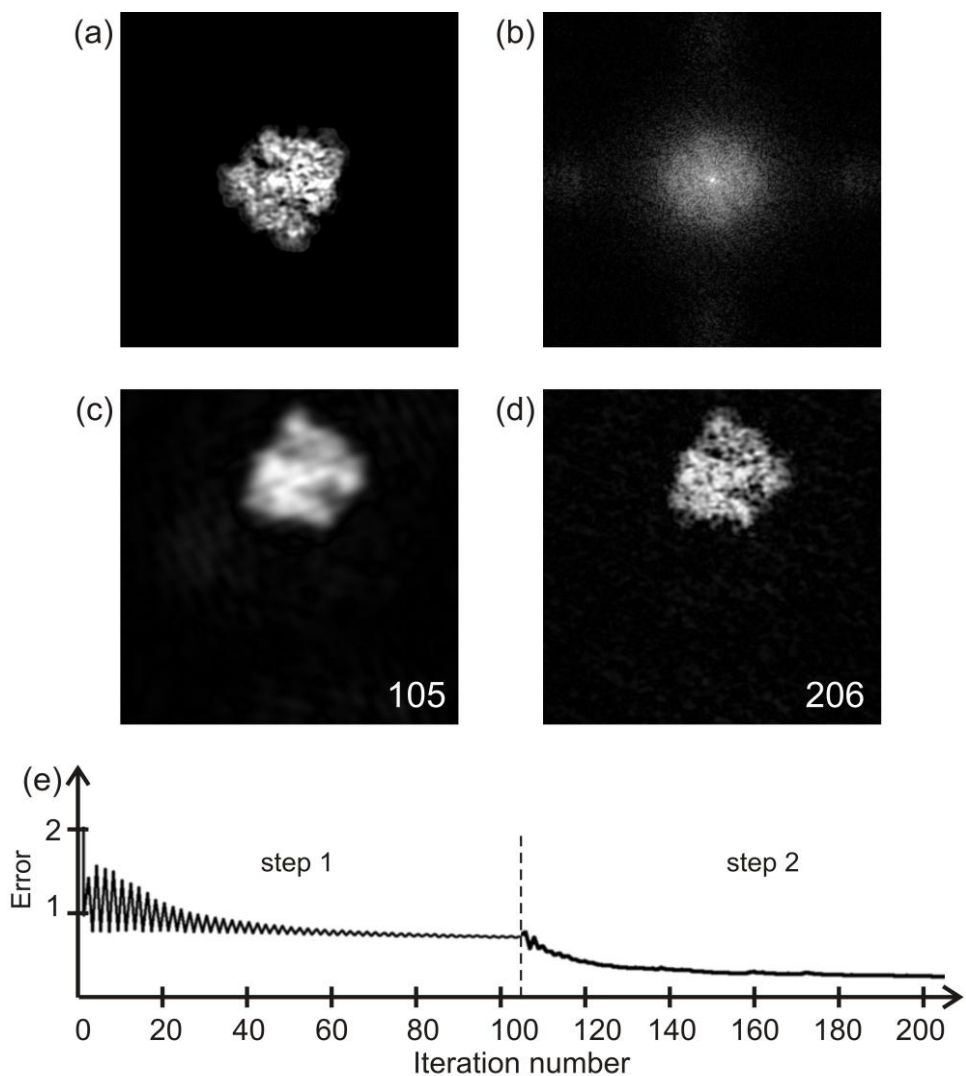

Fig. 4. Two-step recovery of ribosome from its simulated diffraction pattern. (a) Electron density of 70S ribosome. (b) Simulated diffraction pattern shown in logarithmic intensity scale. (c) Amplitude of the low-resolution reconstruction of the object obtained after the first step consisting of 105 iterations with $\mu_1=0.0001$ and $\varepsilon_1=10$. (d) Amplitude of the object after the second step using additional 101 iterations with $\mu_2=1$ and $\varepsilon_2=1000$. (e) Error as a function of iteration number.

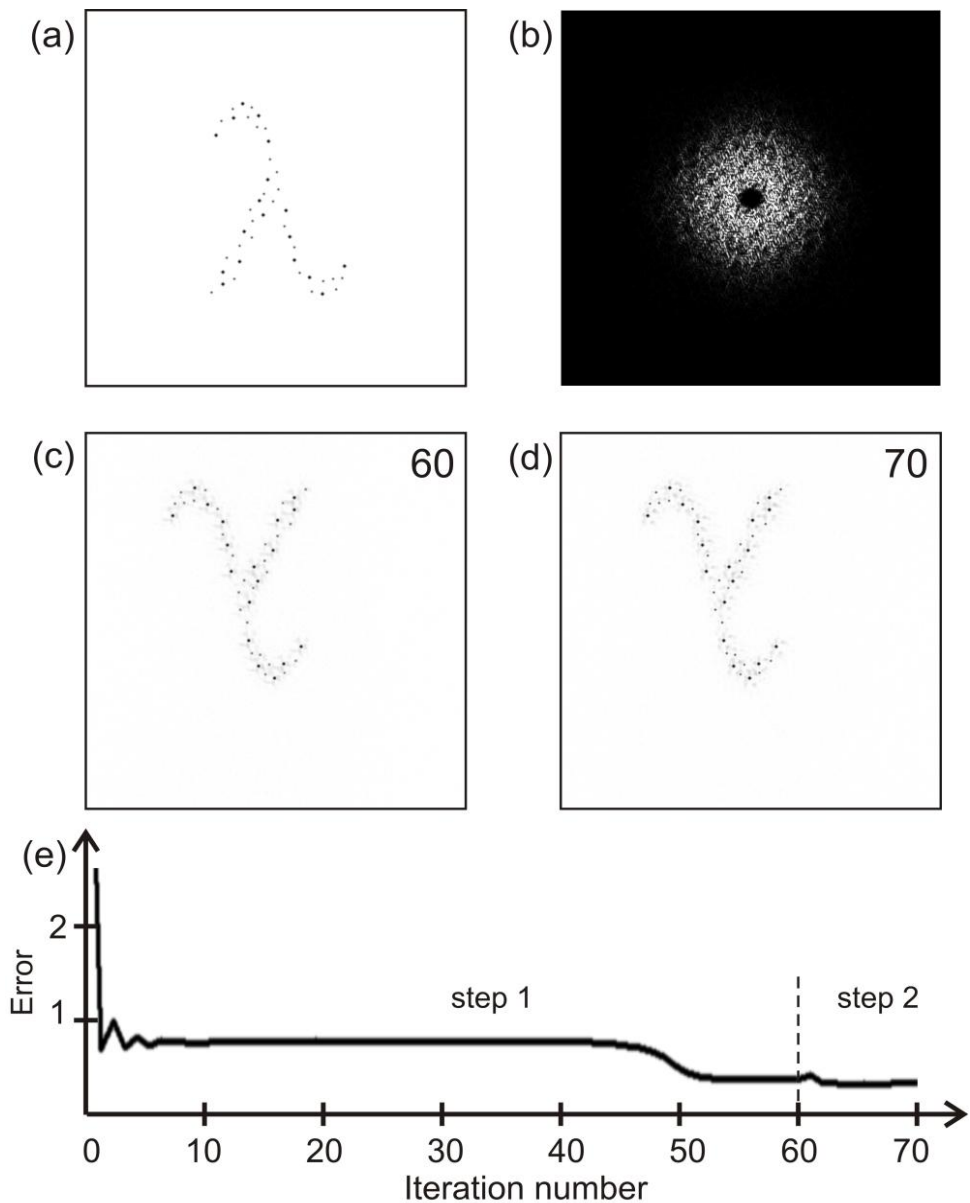

Fig. 5. Two-step recovery of the positions of a set of individual scatterers from its simulated diffraction pattern. (a) Distribution of the scatterers. (b) Simulated diffraction pattern displayed in a logarithmic intensity scale. (c) Amplitude of the low-resolution reconstruction of the objects obtained after the 1st reconstruction step consisting of 60 iterations with $\mu_1=0.01$ and $\varepsilon_1=1$. (d) Amplitude distribution of the objects after the second step using additional 10 iterations with $\mu_2=1$ and $\varepsilon_2=1$. (e) Error as a function of iteration number.

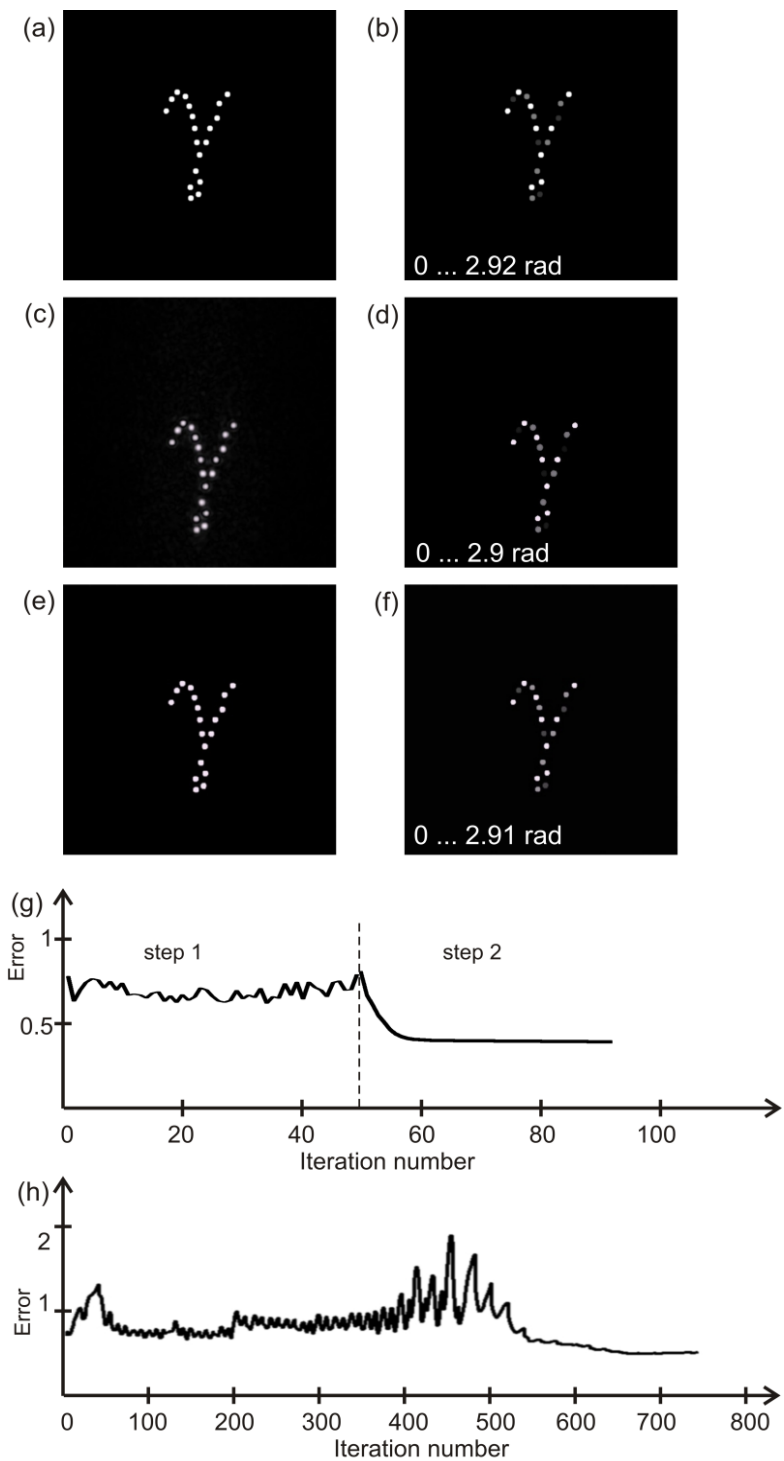

Fig. 6. Recovery of a set of round phase-shifting objects from its simulated diffraction pattern. (a) - (f) Left column: amplitude of the transmission function. Right column: phase of the transmission function. (a) - (b) Original distributions. (c) - (d) Distributions reconstructed by employing our L-filter. (e) - (f) Distributions reconstructed by the shrinkwrap method. (g) Error as a function of iteration number in the reconstruction by the L-filter method. (h) Error as a function of iteration number in the reconstruction by the shrinkwrap method.

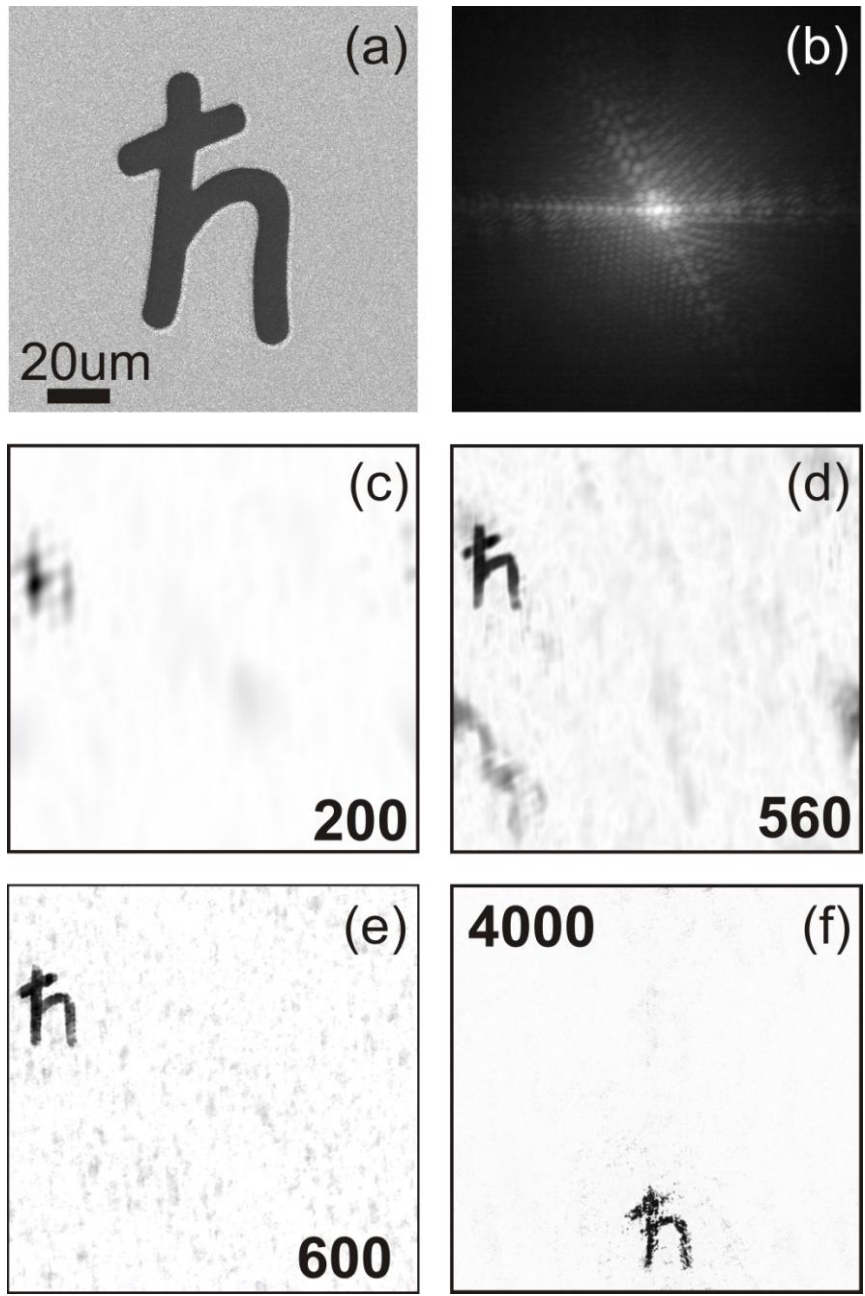

Fig. 7. Recovery of the letter "ℏ" from its experimental diffraction pattern. (a) Scanning electron microscope image of the object. (b) Recorded diffraction pattern displayed in logarithmic intensity scale. (c) Recovered low-resolution image of the object after 200 iterations. (d) Recovered amplitude of the object after 560 iterations, (e) after 600 iterations. (f) Recovered amplitude of the object after 4000 iterations using a different routine that employs a tight mask.